\pdfoutput=1
\documentclass[12pt,preprint]{aastex}
\usepackage[margin=0.75in]{geometry}
\usepackage{graphicx}
\usepackage{natbib}
\usepackage{aas_macros}
\usepackage[section] {placeins}
\usepackage{subfigure}
\usepackage{float}
\usepackage{color}
\usepackage{footnote}

\usepackage{amsmath}

\def\msun{{\rm\,M_\odot}}

\def\msun{{\rm\,M_\odot}}

\newcommand{\kms}{\, {\rm km\, s}^{-1}}

\newcommand{\mpc}{\, {\rm Mpc}}
\newcommand{\kpc}{\, {\rm kpc}}
\newcommand{\hmpc}{\, h^{-1} \mpc}

\newcommand{\hi}{\mbox{H\,{\scriptsize I}\ }}
\newcommand{\heii}{\mbox{He\,{\scriptsize II}\ }}

\def\h2{${\rm\,H_2}$}

\makeatletter 

\def\hi{\noindent \hangindent=2.5em}

\def\kpc{{\rm\,kpc}}

\def\mpc{{\rm\,Mpc}}

\def\kms{{\rm\,km/s}}
\def\msun{{\rm\,M_\odot}}

\def\vol#1  {{{#1}{\rm,}\ }}

\def\hi{{H\thinspace I}\ }

\def\heii{{He\thinspace II}\ }

\newcount\refno
\newcount\rfno
\def\eq{$^{\the\refno\ }$\advance\refno by 1}
\def\ad{\advance\rfno by 1}

\def\clock{\count0=\time \divide\count0 by 60
     \count1=\count0 \multiply\count1 by -60 \advance\count1 by \time
     \number\count0:\ifnum\count1<10{0\number\count1}\else\number\count1\fi}
\def\draft{
   \rightline{DRAFT: \today \quad\quad \clock}}
\def\myputfigure#1#2#3#4#5%
{\hskip0.03\textwidth\vskip#5pt
\makebox[0pt]{\hskip#2in
\includegraphics[width=#3\textwidth]{#1}}\vskip#4pt\hfill}

\newcount\refno
\newcount\rfno
\def\eq{$^{\the\refno\ }$\advance\refno by 1}
\def\ad{\advance\rfno by 1}

\renewcommand{\refname}{Conclusions}

\definecolor{burntorange}{rgb}{1,0.4,0.2}

\begin{document}

\title{Physics of Prodigious Lyman Continuum Leakers}

\author{
Renyue Cen$^{1}$ 
}

\footnotetext[1]{Princeton University Observatory, Princeton, NJ 08544;
 cen@astro.princeton.edu}

\begin{abstract} 

An analysis of the dynamics of a star formation event is performed.
It is shown that galaxies able to drive leftover gas to sufficient altitudes in a few million years
are characterized by two basic properties: small sizes ($\le 1$kpc) 
and high star formation rate surface densities ($\Sigma_{\rm SFR}\ge 10\msun~{\rm yr}^{-1}~{\rm kpc}^{-2}$).
For the parameter space of relevance, the outflow is primarily driven by supernovae
with radiation pressure being significant but subdominant.
Our analysis provides the unifying physical origin 
for a diverse set of observed LyC leakers,
including the green-peas galaxies, [SII]-weak galaxies, Lyman-alpha emitters,
with these two characteristics as the common denominator.
Among verifiable physical properties of LyC leakers, we predict that 
(1) the newly formed stellar masses are are typically in the range of $10^8-10^{10}\msun$,
except perhaps ULIRGs,
(2) the outflow velocities are typically in the range typically of $100-600\kms$, 
but may exceed $10^3\kms$ in ULIRGs,
with a strong positive correlation between the stellar masses formed and the outflow velocities,
(3) the overall escape fraction of galaxies is expected to increase with increasing redshift,
given the cosmological trend that galaxies become denser and more compact with increasing redshift.
In addition, two interesting by-product predictions are also borne out.
First, ULIRGs appear to be in a parameter region where they should be prodigious LyC leakers,
unless there is a large ram-pressure due to infalling gas with a rate exceeding about 30 times
the star formation rate.
Then, towards the tail end of a ULIRG event when the ram-pressure relents, advanced ULIRGs 
are expected to leak more LyC photons than earlier ULIRGs.
Second, Lyman break galaxies (LBGs) are not supposed to be prodigious LyC leakers in our model, 
given their claimed effective radii exceeding $1$kpc.
Thus, if LBGs are observed to have LyC leakers, it may be that the effective radii of their 
star forming regions have been over-estimated by a factor of $2-4$.

\end{abstract}

\section{Introduction}

Understanding how Lyman continuum photons (LyC) escape from galaxies is necessary
for understanding the epoch of reionization (EoR), one of the last major frontiers of astrophysics.
High resolution cosmological hydrodynamic galaxy formation simulations have widely evidenced
that supernova feedback driven blastwaves are the primary facilitator 
to evacuate or create major pores in the interstellar medium to enable the escape of LyC
\citep[e.g.,][]{2009Wise, 2014Kimm, 2015bCen, 2016Ma, 2019Kimm}.
Since LyC escape is not directly measurable at EoR due to its limited mean free path,
it is imperative to ascertain this unknown
by establishing observable proxies for the escape fraction, $f_{esc}$,
when both proxies and $f_{esc}$ are measurable at lower redshift,
based upon a satisfactory physical understanding.

Observationally, in the low-z ($z<0.4$) universe the majority of galaxies with large $f_{esc}$ values turn out to belong to
the compact, so-called green-peas galaxies from the SDSS sample, characterized by their low stellar masses,
low metallicities, very strong nebular emission-lines (H$\beta$ equivalent widths ${\rm > 200}$\AA)
and very high flux ratios of ${\rm [OIII]5007/[OII]3727 > 5}$ \citep[e.g.,][]{2016Schaerer,
2016Izotov, 2016bIzotov, 2018Izotov, 2018bIzotov, 2019Izotov}.
Interestingly, the green-peas galaxies have
star-formation rate surface densities of ${\rm 10-100 \msun yr^{-1} kpc^{-2}}$,
which are much higher than typical star-forming galaxies in the
local universe but may be similar to those at EoR.
Another class of low redshift galaxies that have high LyC escape fraction 
is identified by their high Ly$\alpha$ emission \citep[e.g.,][]{2015Verhamme, 2017Verhamme},
which typically have star-formation rate surface densities of ${\rm \sim 10 \msun yr^{-1} kpc^{-2}}$.
At $z\sim 3$  
LyC escape is detected in dozens of individual galaxies
\citep[e.g.,][]{2015Mostardi, 2016Vanzella, 2016Shapley, 2018Steidel},
some of which also show intense [OIII] emission that are consistent with low-z observations
and characteristic of galaxies at EoR \citep[e.g.,][]{2019Fletcher}.
Furthermore, recently, another set of galaxies with relatively weak 
[SII] nebular emission lines are also observed to show high LyC escape \citep[][]{2019Wang}.
The low-redshift green-peas galaxies, $z\sim 3$ high LyC leakers
and the [SII]-weak LyC leaking galaxies are different 
in various respects, such as stellar mass, metallicity, dust content and ISM properties.
But all appear to share two common characteristics:
all four have very high star-formation rate surface densities and relatively compact sizes.

This {\it Letter} aims to understand 
if supernova feedback may be the common physical process 
that underwrites the commonality shared by these different classes of galaxies observed.
We will show that this is indeed the case.
This finding thus provides a physical basis to help identify galaxies with high LyC leakage at the epoch of reionization 
by indirect but robust markers that can be established at more accessible redshift and
for why dwarf galaxies at EoR are much more capable of enabling high LyC escape fraction than typical low redshift counterparts.

\section{Physics of Lyman Continuum Leakers}

We explore if gas density-bound structures in star forming galaxies may be produced.
The following treatment is undoubtedly simplified but capture the essence of the physics,
and is primarily a means to identify likely physical parameter space that is relevant for making galaxies with high LyC escape fractions.
A gas cloud of initial mass $M_{gas,0}$ with a half light radius $r_h$ 
and a star formation rate SFR gives rise to an outward radial force on the gas cloud itself due to supernova explosion generated radial 
momentum equal to 
\begin{equation}
\label{eq:balance}
  \begin{split}
F_{SN} = SFR \times p_{SN} \times M_{SN}^{-1},
  \end{split}
\end{equation}
\noindent
where $p_{SN}=3\times 10^5\msun\kms$ is the terminal momentum generated per supernova \citep[e.g.,][]{2014Kimm},
$M_{SN}$ is the amount of stellar mass formed to produce one supernova, which is equal to about $(50,75,100)\msun$
for (Chabrier, Kroupa, Salpeter) initial mass function (IMF), respectively.
The exact value of $p_{SN}$ weakly depends on density and metallicity of the ambient gas.
For simplicity without loss of validity given the concerned precision of our treatment,
we use the above fiducial value.
Another mechanical form of feedback from massive stars is fast stellar winds due to O stars.
The total energy from stellar winds is about a factor of ten lower than the total energy from supernovae
\citep[e.g.,][]{1999Leitherer}. Since stellar winds roughly 
track core collapse supernovae, we simply omit stellar winds bearing a loss of accuracy at 10\% level.
The second important outward force on the gas is the radiation pressure on dust grains, equal to 
\begin{equation}
\label{eq:frad}
  \begin{split}
F_{rad} = SFR \times \alpha \times c \left[1-\exp{(-{\Sigma_{gas}\kappa_{UV}})}\right](1+{\Sigma_{gas}\kappa_{FIR})},
  \end{split}
\end{equation}
\noindent
where $\alpha=3.6\times 10^{-4}$ is an adopted nuclear synthesis energy conversion efficiency from rest mass to radiation, $c$ speed of light,
$\kappa_{UV}=1800 cm^2~g^{-1}$ 
and $\kappa_{FIR}=20 cm^2~g^{-1}$
the opacity at UV \citep[e.g.,][]{2003Draine} and 
dust processed radiation far infrared (FIR) radiation \citep[e.g.,][]{2017Lenz}, respectively,
$\Sigma_{gas}$ the surface density of the gas.
The exact value of 
$\kappa_{UV}$ matters little in the regime of interest but variations of the value of $\kappa_{FIR}$ does matter to some extent.
To place the two forces in relative terms,
we note that at ${\rm \Sigma_{gas}=1.3\times 10^4\msun~pc^{-2}}$,
the radiation pressure due to IR photons equals the ram pressure due to supernova blastwaves,
with the former and latter dominating at the higher and lower surface densities, respectively.

There are two relevant inward forces.
The mean gravitational force, when averaged over an isothermal sphere, which is assumed, is 
\begin{equation}
\label{eq:Fg}
  \begin{split}
F_g = {\ln (r_{max}/r_{min}) GM_{gas,0}M_{gas}(t)\over 4 r_h^2},
  \end{split}
\end{equation}
\noindent
where $r_{min}$ to $r_{max}$ are minimum and maximum radii of the gas cloud being expelled.
For our calculations below we adopt $r_{min}=100$pc and $r_{max}=r_h$;
the results depend weakly on the particular choices of these two radii.
We note that $M_{gas}(t)$ is the remaining mass of the gas cloud when it starts to be lifted at time $t_L$ by the combined
force of supernova driven momentum flux and radiation momentum flux against inward forces,
with $M_{gas,0}-M_{gas}(t_L)$ having formed into stars. 
Another inward force is that due to ram pressure, which we parameterize in terms of gas infall rate in units
of star formation rate:
\begin{equation}
\label{eq:Frp}
  \begin{split}
F_{rp} = \dot M_{inf} v_{inf} = \eta \times SFR\times \left({GM_{gas,0}\over r_h}\right)^{1/2},
  \end{split}
\end{equation}
\noindent
where $\eta$ is the ratio of mass infall rate to SFR.

The relevant physical regime in hand is how to drive the gas by the combined force
of supernovae and radiation against the combined force of gravitational force and ram-pressure.
A key physical requirement, we propose,
is that the feedback process needs to promptly lift the entire remaining gas cloud to a sufficient height
such that it piles itself into a (thin) shell that subsequently fragments while continuing moving out,
in order to make a copious LyC leaker.
It seems appropriate to define ``a sufficient height" 
as a height on the order of $r_h$, which we simplify to be just $r_h$.
The above definition may be expressed as
\begin{equation}
\label{eq:drive}
  \begin{split}
(F_{SN} + F_{rad} - F_g - F_{rp}) (t_h-t_L) = M_{gas}(t_L) v_h \quad\quad {\rm and} \quad\quad 2 r_h = (t_h-t_L) v_h,
  \end{split}
\end{equation}
\noindent
where $v_h$ is the shell velocity when reaching $r_h$ at time $t_h$,
and $M_{gas}(t_L)$ is the gas cloud mass at $t_L$ when it begins its ascent.

\begin{figure}[!h]
\centering
\vskip -0.0cm
\resizebox{6.0in}{!}{\includegraphics[angle=0]{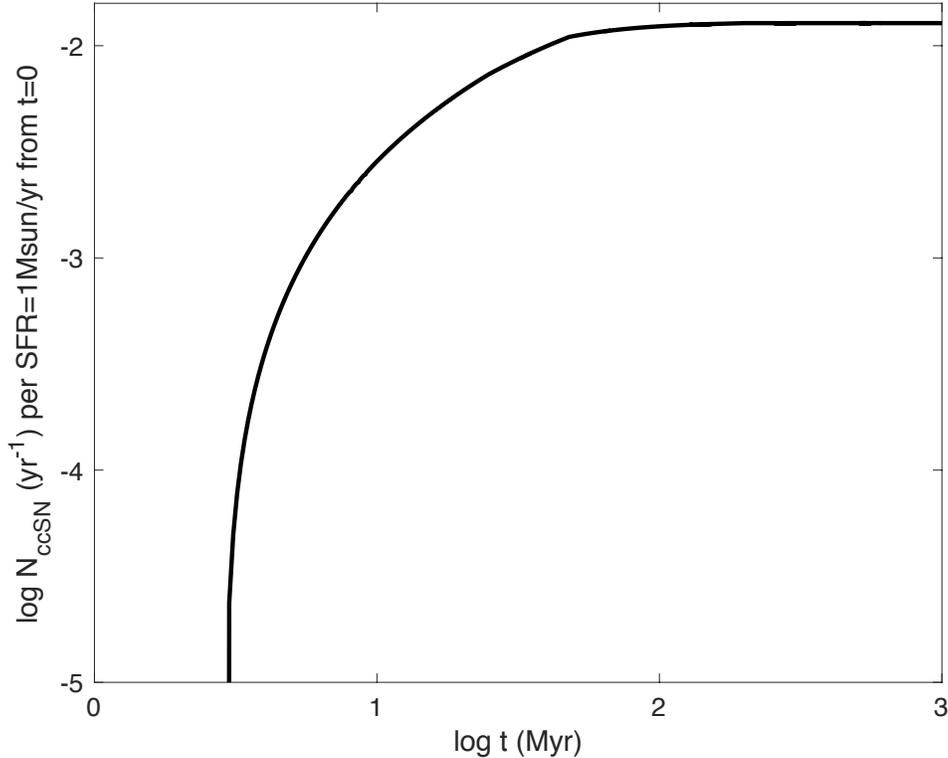}}
\vskip 0.0cm
\caption{
shows the supernova rate for a star formation event at a star formation rate of $1 \msun/yr$ starting at time $t=0$ as a function of time.
This plot is produced using Eq (A.2) of \citet[][]{2017Zapartas} of the core-collapse supernova rate including both single stars
and binary mergers.
We note that at $t\ge 200$Myr the saturation rate corresponds to one supernova per 78$\msun$ of stars formed,
approximately in agreement with what a Kroupa IMF gives.
}
\label{fig:ramp}
\end{figure}

We may relate the initial gas mass $M_{gas,0}$ to SFR that is observable by using an empirically found relation:
\begin{equation}
\label{eq:SFR}
  \begin{split}
SFR = c_* M_{gas,0}/t_{dyn},
  \end{split}
\end{equation}
\noindent
where $G$ is the gravitational constant,
$t_{dyn} = \sqrt{{3\pi\over 32G\rho_t}}$ is the dynamical time of the system
with $\rho_t$ being the total density, the sum of gas and stars within $r_h$,
star formation efficiency per dynamical time is found to be $c_*=0.01$ \citep[][]{2012Krumholz}.
Note that the SFR above is the SFR up to the time $t_L$, when it is shut down upon the uplift of the gas cloud.

We compute the rate of supernova explosion more precisely.
This is needed because as soon as the combined outward force of supernova feedback and radiation pressure
is stronger than inward forces at time $t_L$, we need to stop star formation then.
It is possible in some cases that the star formation has not lasted long enough to reach the saturation supernova rate.
We use a recent, comprehensive analysis of \citet[][]{2017Zapartas} that
takes into account both single and binary stellar populations, including supernovae due to binary mergers.
We convolve the fitting formula (A.2) in \citet[][]{2017Zapartas} that 
is composed of three separate temporal segments, $3-25$Myr and $25-48$Myr due to massive single stars
and $448-200$Myr due to binary merger produced core-collapse supernovae,
with a constant star formation rate SFR (Eq \ref{eq:SFR}) starting at time $t=0$.
Figure (\ref{fig:ramp}) shows the resulting instantaneous
supernova rate as a function of time for a star formation event at a constant $SFR=1 \msun/yr$.
Then, $M_{SN}(t) = 1\msun/N_{ccSN}$ (where $N_{ccSN}$ is the y-axis shown in Figure \ref{fig:ramp})
as a function of time since the start of the starburst,
in lieu of a constant value of $M_{SN}$ that is the saturation value at $t\ge 200$Mpr,
in Eq (\ref{eq:balance}), where appropriate.
We note that at $t\ge 200$Myr the saturation rate corresponds to one supernova per 78$\msun$ of stars formed,
approximately corresponding to a Kroupa IMF.

\begin{figure}[!h]
\centering
\vskip -1.0cm
\resizebox{6.0in}{!}{\includegraphics[angle=0]{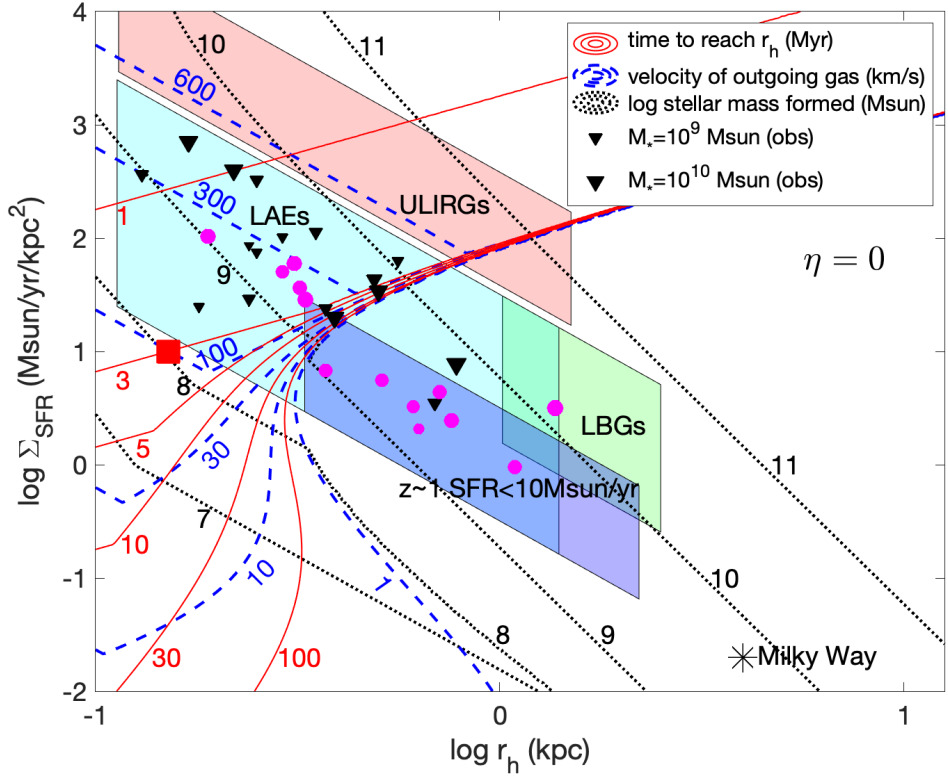}}
\vskip 0.0cm
\caption{
shows the time that it takes to evacuate the gas to an altitude of $r_h$, $t_h-t_L$,
as the solid red contours labelled in units of Myr,
with labels ``1", ``3", ``5", ``10", ``30" and ``100".
Shown as dotted black contours are the log of the stellar mass in units of $\msun$ formed from this episode,
with labels ``8", ``9", ``10" and ``11".
The dashed blue contours depict the radial velocity of gas being lifted in units of $\kms$, 
with labels ``10", ``30", ``100", ``300" and ``600".
The shaded light blue, light green, light red and dark blue regions indicate approximately regions normally 
referred to Lyman alpha emitters (LAEs), 
Lyman break galaxies (LBGs) at high redshift,
ultra luminous infrared galaxies (ULIRGs), 
and $z\sim 1$ star-forming but non-LyC leaking dwarf galaxies,
respectively.
The LAE region is obtained by using a radius range of $0.1-1.4$kpc
and a range of SFR of $1-100{\rm \msun~yr^{-1}}$ \citep[e.g.,][]{2007Gawiser,2009Bond}.
The LBG region is obtained by using a radius range of $1.2-2.5$kpc
and a range of SFR of $5-100{\rm \msun~yr^{-1}}$ \citep[][]{2002Giavalisco}.
The ULIRG region is approximately delineated by 
a radius range of $0.1-1.5$kpc
and a range of SFR of $120-1200{\rm \msun~yr^{-1}}$ \citep[e.g.,][]{2018Spence}.
The location of the Milky Way galaxy is indicated by a black star near the lower-right corner.
The sample of star-forming but non-LyC leaking dwarf galaxies at $z\sim 1$ with $SFR < 10\msun~yr^{-1}$
\citep[][]{2016Rutkowski} is the blue shaded region labelled as ``$z\sim 1$ SFR $<10\msun$/yr".
Finally, the observed galaxies with large LyC escape fractions are shown as 
black downward-pointing triangles from various sources
\citep[e.g.,][]{2015Alexandroff, 2016Izotov, 2016bIzotov, 2018Izotov, 2018bIzotov, 2019Wang},
where some galaxies known as LAEs but with little LyC escape are shown as solid magenta dots \citep[][]{2015Alexandroff}.
In all cases for $M_*$, $r_h$ and SFR of observed LyC leakers and non-leakers
we use updated values from \citet[][]{2019Wang}.
}
\label{fig:sol3v}
\end{figure}

Figure (\ref{fig:sol3v}) shows the results by integrating Eq (\ref{eq:drive}).
The solid red contours labelled in units of Myr shows the time, $t_h-t_L$, which it takes to drive the gas to an altitude of $r_h$.
Earlier we have mentioned the need of ``promptly" driving the gas away, which we now elaborate.
For any starburst event, massive O stars formed that dominate the LyC radiation die in about 5Myr.
Therefore, the time elapsed since the end of the starburst of the observed prodigious LyC leakers should not be longer
than that time scale, i.e., $t_h-t_L\le 5 Myr$.
Comparing the $t_h-t_L=5$Myr contour with the black solid triangles 
indicates that all the observed LyC leakers lie in the parameter region with 
$t_h-t_L\le 5Myr$, except J0921 with $r_h=0.78$kpc, $SFR=7.68\msun~yr^{-1}$ and $M_*=6.3\times 10^{10}\msun$ 
and J0926 with $r_h=0.69$kpc, $SFR=3.47\msun~yr^{-1}$ and $M_*=1.3\times 10^{9}\msun$ \citep[][]{2015Alexandroff}.
In the entire region of possible prodigious LyC leakers we point out that
the outward force is dominated by supernova driven momentum, although in a thin top-left wedge region the radiation pressure alone
is also able to counter the gravity.
It is very clear that all LyC leakers live in a parameter space generally denoted as Lyman alpha emitters,
as indicated by the large, cyan-shaded region \citep[e.g.,][]{2007Gawiser,2009Bond}.
However, it is also clear that not all LAEs are LyC leakers, as noted by the magenta dots that are
observed to be LyC non-leakers.
We interpret this as that the gas being lifted by the supernova driven momentum
is fragmentary such that obscuration or transparency of the LyC sources are sightline dependent even when the gas cloud as a whole 
is expelled to a high altitude.
We note that, if one includes binary evolution effects, such as merger produced blue stragglers or stripped hot
helium stars, additional O stars like stars will emerge with some delay of order $10$Myr.
Each of these two delayed components may mount to about 10\% of LyC photons produced by initial starburst
\citep[][]{2017Eldridge}.
This may be a significant addition of LyC sources.
Nevertheless, given the closely spaced red contours in 
Figure (\ref{fig:sol3v}), we see that none of our conclusions will be significantly altered,
if we use $t_h-t_L=10Myr$ instead of $t_h-t_L=5Myr$.

In the right-side, large gulf region occupying about one half of the plot area,
gravity dominates over the combined outward force of supernova explosion driven momentum flux and radiation pressure.
In this region, no complete lift-up of gas to $r_h$ is possible regardless of the duration of the star formation episode.
This region contain the blue shaded region labelled as ``$z\sim 1$ SFR $<10\msun$/yr",
which is a sample of star-forming dwarf galaxies at $z\sim 1$ with $SFR < 10\msun~yr^{-1}$
that do not show significant LyC leakage \citep[][]{2016Rutkowski}.
The fact that this region lies in the region of the parameter space 
that is part of the LAE region and indeed is expected not to have large LyC escape is quite remarkable,
because the author was not aware of this data set until was brought attention to it by the referee.
Also in the large gulf region are the LBGs, as indicated by the green-shaded region \citep[][]{2002Giavalisco},
suggesting that LBGs are not likely to be copious LyC leakers.
However, recent observations \citep[][]{2018Steidel}
indicate a mean $f_{esc}=0.09\pm 0.01$ for a subsample of LBGs.
This directly contradicts our conclusions.
One possible way of reconciliation is that the observed effective radii
of LBGs in UV may be over-estimates of the effective radii of the star-forming regions;
if the actual radii of star-forming regions are in the range of $300-500$pc,
LBGs would be located in the region of LyC leakers.
Alternatively, star-forming regions of LBGs may be composed of much more compact sub-regions.
While not direct proof, it is intriguing to note that
\citet[][]{2009Overzier} find that the three brightest of their sample of thirty galaxies
low-redshift analogs of LBGs at $z=0.1–0.3$ that they examine in detail  
indeed have very compact sizes, with effective radii no larger than $70-160$ pc.
Thus, it would be significant to carry out high resolution FIR observations, such as by ALMA,
of LBGs to verify if the total star-forming regions are in fact more compact.

As another example, the star near the bottom-right is the location of the Milky Way, which is also inside 
the LyC non-leaker region.
So our Galaxy is unlikely to be a very good LyC leaker for an extragalactic observer.
On the other hand, a class of very luminous galaxies - ULIRGs - occupies a region
that may straddle the LyC leaker and non-leaker region.
ULIRGs are in a special region of the parameter space.
It is known that ULIRGs are copious FIR emitters, not known to be LyC leakers.
We suggest that ULIRGs may belong to a class of its own, where ram-pressure due to gas infall 
may have helped confine the gas to (1) make them LyC non-leakers and 
(2) allow for star formation to proceed over a much longer period than
indicated by the red contours, despite the strong outward momentum flux driven by ongoing star formation.
One way to test this scenario is to search for redshifted 21~cm absorption lines in ULIRGs,
if suitable background radio quasars/galaxies or intrinsic central radio quasars/galaxies or possible other bright radio sources.

\begin{figure}[!h]
\centering
\vskip -0.0cm
\resizebox{5.4in}{!}{\includegraphics[angle=0]{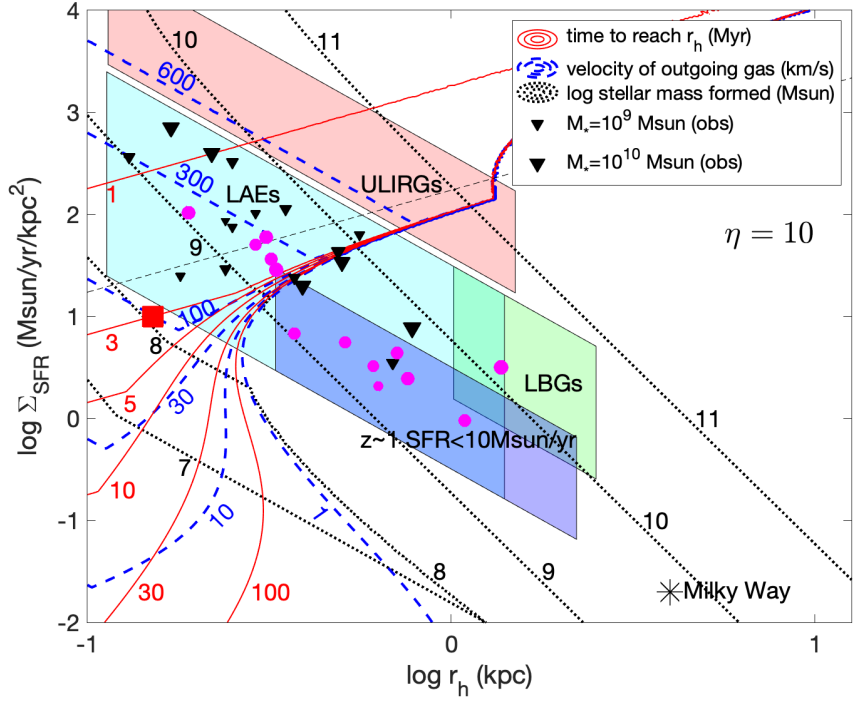}}
\resizebox{5.4in}{!}{\includegraphics[angle=0]{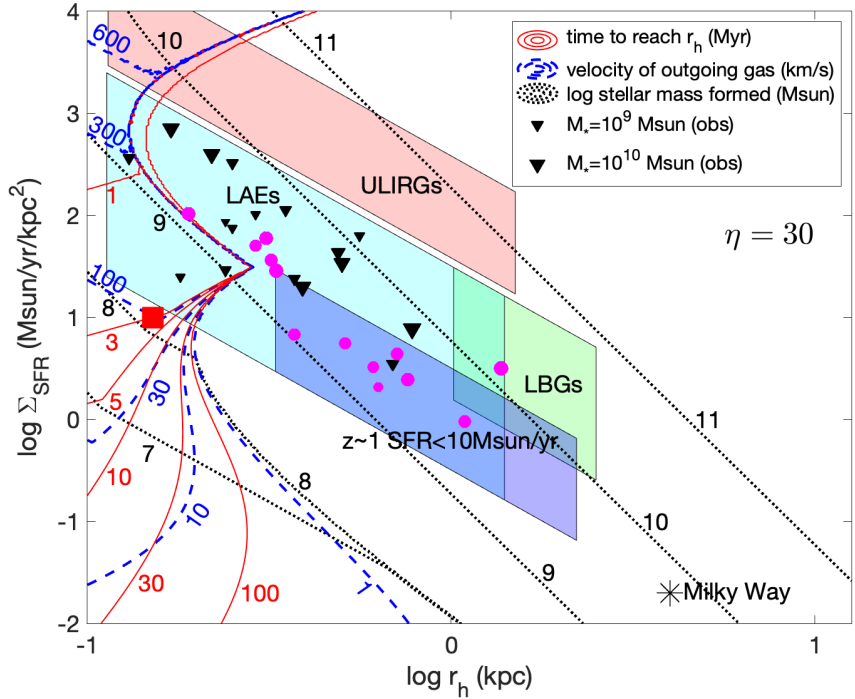}}
\vskip 0.0cm
\caption{
Top panel is similar to Figure (\ref{fig:sol3v}) with one change: $\eta=10$ is used here instead of $\eta=0$ (see Eq \ref{eq:Frp}) 
in Figure (\ref{fig:sol3v}).
Bottom panel is similar to the top panel with $\eta=30$.
}
\label{fig:sol3v2}
\end{figure}

Nevertheless, we would like to point out that ULIRGs should vary as well.
Imagine a merger or other significant event drives an episode 
of cold gas inflow. The episode spans a period and the starburst triggered
goes from the initial phase of buildup when the star formation rate 
is extremely subdominant to the inflow gas rate. 
An estimate of possible gas inflow rates is in order to illustrate the physical plausibility of this scenario.
Let us assume that a merger of two galaxies each of halo mass of $10^{12}\msun$ and gas mass $1.6\times 10^{11}\msun$
triggers a ULIRG event and 
that 10\% of the total gas mass falling onto the central region of size $1$kpc at a velocity of $300\kms$.
Then we obtain a gas infall rate of $\dot M_{in} = 1.0\times 10^{4}\msun~{\rm yr}^{-1}$, which would correspond
yield $\eta=(100,10)$ for SFR equal to $(100,1000)\msun~{\rm yr}^{-1}$, respectively.
In Figure (\ref{fig:sol3v2}) we see that, once the infall rate drops below about $30$ times
the SFR, gas in ULIRGs would be lifted up by supernovae.
This leads to a maximum SFR in ULIRGs that is estimated as follows using this specific merger example.
During the buildup phase of the ULIRG, since the gas infall rate exceeds greatly the SFR,
one can equate the gas mass to the total dynamical mass.
Thus, we have ${\rm SFR}=c_* M_{gas} [r/(GM_{gas}/r)^{1/2}]^{-1}$.
Equating $\eta {\rm SFR}$ (with $\eta=30$) to $\dot M_{in}$, we find 
the amount of gas accumulated at the maximum gas mass 
is $M_{gas,max}=6.3\times 10^{10}\msun$, corresponding to a maximum SFR $SFR_{max}=330\msun~{\rm yr}^{-1}$ in this case.
Thus, our analysis indicates that the physical reason for an apparent
maximum SFR in ULIRGs and SMGs may be due to a competition between the maximum ram-pressure confinement of gas
and internal supernovae blastwave and radiation pressure.
This contrasts with and calls into question
the conventional view of radiation-pressure alone induced limit on maximum SFR \citep[e.g.,][]{2005Thompson}.
We deferred a more detailed analysis on this subject to a separate paper.

At a later point in time it may be transitioned sufficiently rapidly, at least for a subset
of ULIRGs, to a phase that is ubiquitous in outflows.
Some ULIRGs at this later phase may become significant LyC leakers,
if and when the gas inflow rate drops below about 10 times SFR, as shown in 
Figure (\ref{fig:sol3v2}) by varying the ram-pressure (the $\eta$ parameter, see Eq \ref{eq:Frp}).
This new prediction is in fact
consistent with some observational evidence that shows 
significant Ly$\alpha$ and possibly LyC escape fractions in the advanced stages
of ULIRGs \citep[e.g.,][]{2015Martin}.
These ULIRGs also seem to show blueshifted outflow.
It ought to be noted that their measured $f_{esc}$ is relative to the observed FUV luminosity
(i.e., the unobscured region)
but not relative to the total SFR, which is difficult to measure.
Thus, the escape of LyC in these late stage ULIRGs is a relative statement 
compared to ULIRGs that are ram-pressure confined and are not LyC leakers
in the sense that, although in the former the stellar feedback processes may be 
able to lift gas up, likely still substantial inflow gas may be able to 
continue to provide a large amount of obscuring material, 
albeit less than at earlier phase with a stronger ram-pressure confinement and heavier obscuration.

Let us now turn to the black dotted contours showing 
the log of the stellar mass in units of $\msun$ formed from this episode presumably triggered by a gas accretion event.
Two points are worth noting here.
First, in the region where LyC leakers are observed,
the expected stellar mass formed in a single star formation episode 
is in the range of $10^8-10^{10}\msun$.
The observed green-peas galaxies \citep[e.g.,][]{2016Schaerer,
2016Izotov, 2016bIzotov, 2018Izotov, 2018bIzotov, 2019Izotov} have stellar masses
indeed falling in this range.
This suggests that a large fraction or all of the stars in green-peas galaxies may be formed in this most recent
star formation episode.
However, some of the [SII]-weak selected galaxies have stellar masses significantly exceeding $10^{10}\msun$ 
\citep[][]{2019Wang}.
We suggest that in those cases a large fraction of the stars are formed in previous star formation episodes and
spatially more extended than the most recent episode.
In both cases - green-peas galaxies and [SII]-weak galaxies -
given the central concentration of this most recent star formation episode,
it is likely triggered by a low-angular momentum gas inflow event.
It would be rewarding to searches for signs of such a triggering event,
such as nearby companions or post merger features.
Second, there are discontinuities of the contour lines going from the gravity-dominated lower-right region to
the outward force dominated upper-left region.
This is because, while the gas forms to stars unimpeded in the former,
a portion of the gas is blown away in the latter.

\begin{figure}[!h]
\centering
\vskip -0.0cm
\resizebox{5.4in}{!}{\includegraphics[angle=0]{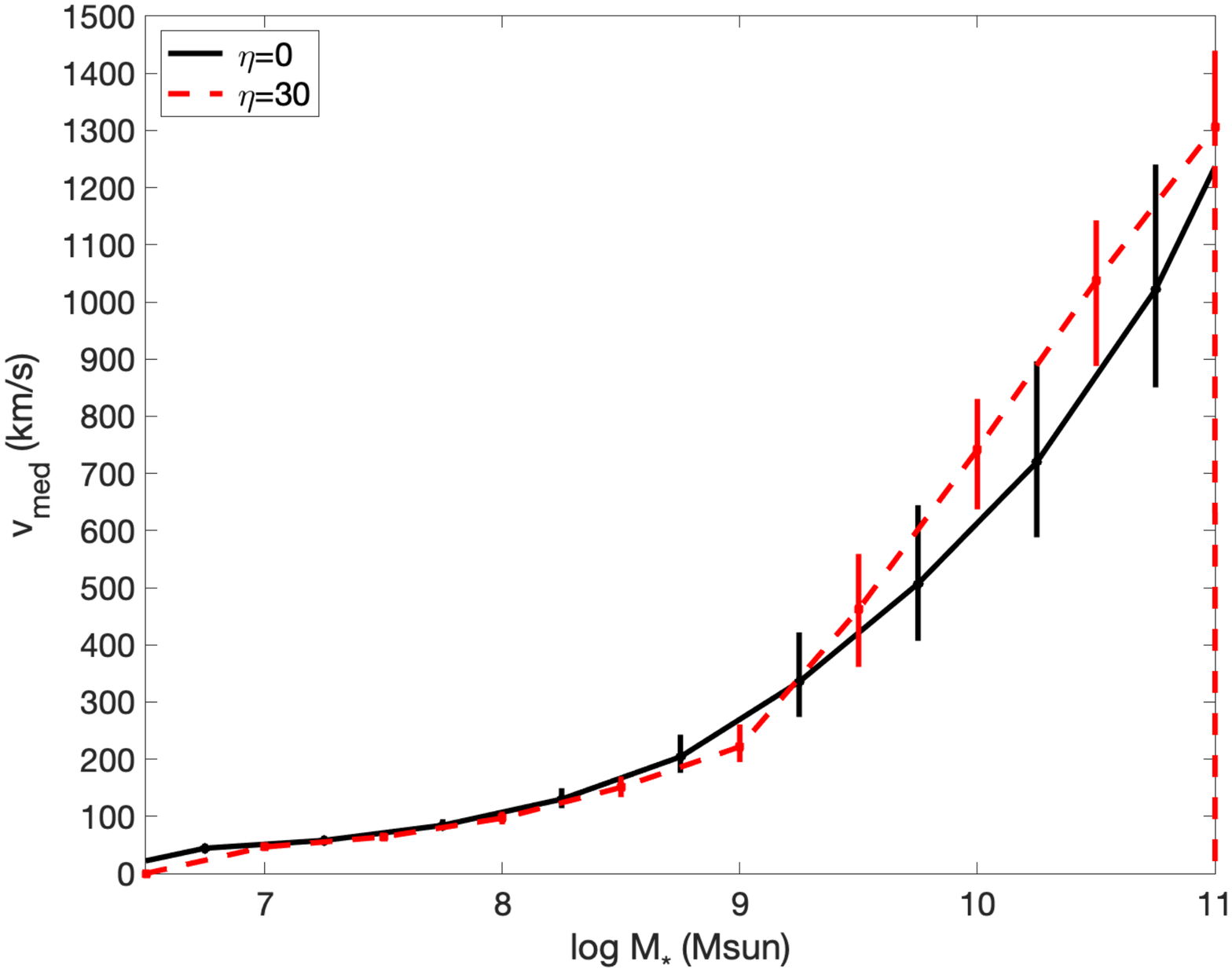}}
\vskip 0.0cm
\caption{
shows the median velocity as a function of stellar mass formed in the episode, 
along with lower and upper quartiles shown as the errorbars,
for two cases with $\eta=0$ and $\eta=30$.
}
\label{fig:vMstar}
\end{figure}

Finally, let us turn our attention to the velocity of the gas moving out, as indicated
by the blue dashed contours.
We see that the outward velocity is in the range of $100-600\kms$.
This is a prediction that can be 
verified by observations when a reasonably large set of data becomes available.
Worth noting is that LyC leakers do not necessarily possess outsized outflow velocities.
At the present time, the sample of LyC leakers is still relative small 
but the approximate range of wind speeds in the range of 
$150-420\kms$ if one uses directly the separation of Ly$\alpha$ peaks as a proxy \citep[][]{2016bIzotov, 2016Izotov}.
We note that given the scattering effects of Ly$\alpha$ photons
the separation of Ly$\alpha$ peaks generally may only represent an upper limit on the velocity dispersion,
which in turn may be on the same order of the outflow velocity.
For a general comparison to young star-forming galaxies without considering LyC escape,
\citet[][]{2013Bradshaw} find outflow velocities typically in the range of $0-650\kms$
for young star-forming galaxies with stellar mass of $\sim 10^{9.5}\msun$,
which is consistent with predicted velocity range.
Finally, 
\citet[][]{2017Chisholm} find that LyC leakers (with $f_{esc}\ge 5\%$)  
spans an outflow velocity range of $50-500\kms$ (probed by Si II),
consistent with our model.
\citet[][]{2015Henry} show outflow velocities
probed by a variety of ions from Si II to Si IV of a range of
$50-550\kms$ for green pea galaxies,
consistent with our model once again.

Because the velocity contours are more parallel than perpendicular to the stellar mass contours,
a related prediction is that the outflow velocity is expected to be positively correlated with
the newly formed stellar mass. 
Figure (\ref{fig:vMstar}) shows 
the median velocity as a function of stellar mass formed in the episode, 
along with lower and upper quartiles shown as the errorbars,
for two cases with $\eta=0$ and $\eta=30$.
We see clearly a positive correlation between the outflow velocity of LyC leakers
and the amount of stars formed in the episode,
with median velocity going from $\sim 100\kms$ at $10^8\msun$ to 
$600-700\kms$ at $10^{10}\msun$.
For the very high end of the stellar mass of $10^{11}\msun$ formed in the episode,
the outflow velocities are expected to exceed $10^3\kms$.
With more data this unique prediction should be testable.

\section{Comparisons to Some Previous Works}

We thank the author for the very detailed comparison with the work I suggested. Since the reader might be unfamiliar with the details of the work which spans observations, simulations, and semi-analytical techniques, I suggest an introductory sentence for each paragraph.

\citet[][]{2001Heckman} are among the first to attempt to infer the physical conditions of LyC escape 
in starburst galaxies combining observational evidence with basic physical considerations in
the context of a superbubble driven by supernova explosions. 
They propose that strong starbursts clear channels through the neutral ISM to facilitate
LyC escape. They ultimately reach the conclusion that the empirical evidence does not
demonstrate that galactic winds inevitably produce large
values of LyC escape fraction in local starbursts. 
In other words, galactic outflows appear to be a necessary but not sufficient condition that
creates an ISM porous to ionizing radiation.
This idea is advanced here in a quantitative fashion.
We show that only very compact, high surface density starbursting regions
are capable of evacuating embedding gas sufficiently promptly to 
allow for an environment where a significant amount of LyC escape becomes possible.
We argue that this may apply to both a compact starburst at the center of a galaxy
or a high density patch of a spatially extended starburst,
because the dynamics are the same in both cases.
Nevertheless, we agree with \citet[][]{2001Heckman} 
that even in this case the condition created by compact strong starbursts
may be a necessary one, due to variations of obscurations along lines of sight,
because in most cases gas is only lifted to a limited altitude 
forming a gas shell that is presumably prone to fragmentation.

In a semi-analytic treatment of escape fraction as a function of star formation surface density,
applied to the Eagle simulation,
\citet[][]{2016Sharma} adopt a threshold star formation surface density $\Sigma_{SFR}=0.1 {\rm \msun~yr^{-1}~kpc^{-2}}$
on a scale of $\sim 1$kpc, motivated by an apparent threshold for driving galactic winds.
Our analysis shows that on $1$kpc scale, such a star formation surface density 
falls short by a factor of 1000 for making conditions to allow for a high LyC escape fraction
(see Figure \ref{fig:sol3v}).
However, when one moves to a smaller size of $0.5$kpc, this 
threshold star formation surface density lands in the region where gas may be driven away but
on a time scale much longer than $5$Myr.
In fact, at $\Sigma_{SFR}=0.1{\rm \msun~yr^{-1}~kpc^{-2}}$ there is no parameter space for
a high LyC escape fraction regardless of size.
For a star formation surface density $\Sigma_{SFR}=1{\rm \msun~yr^{-1}~kpc^{-2}}$,
a region of size $\sim 0.1$kpc can now possess necessary conditions for a high LyC escape fraction.
Thus, the overall LyC escape fraction in Eagle simulation they analyze
may have been over-estimated.
On the other hand, limited numerical resolution may have caused an underestimation of
the star formation surface density in the simulated galaxies there.
Thus, the overall net effect is unclear, 
if all galaxies were resolved and a correct threshold star formation surface density applied.
What is likely is that their assessment of the relative contributions of large and small galaxies
may have been significantly biased for large ones due to the lenient condition.

Based on an empirical model introduced in \citep[][]{2018Tacchella} that
stipulates the SFR to be dependent on halo accretion rate with a redshift-independent star
formation efficiency calibrated by N-body simulations,
\citet[][]{2019Naidu} analyze how observations of electron scattering optical depth and IGM ionization states
may be used to constrain cosmological reionization.
Their main assumption is that the LyC escape fraction is constant for all galaxies.
Their main conclusion is that bright galaxies ($M_{UV}<-16$) are primarily responsible for 
producing most of the ionizing photons, in order to produce a rapid reionization process consistent
with observations.
Our analysis indicates that the assumption of a constant LyC escape fraction for all galaxies
may be far from being correct.
However, if the bright galaxies are dominated by strong compact central starbursts with high star formation surface densities,
an assumed constant LyC escape fraction for all galaxies
may lead to a conclusion, as they do, that faint galaxies make minor contribution to reionization;
this conclusion itself ultimately may not be incorrect, though.
It is also worth noting that the galaxy luminosity in their model 
is substantially shallower than observations below $M_{UV}>-18$.
This discrepancy may have, in part, contributed to the more diminished role of faint galaxies in their modeling.
These coupled effects suggest an improved, more detailed analysis may be desirable,
to better learn the intricate physics.

The dynamics for a central starburst analyzed here 
in principle is applicable to a compact starbursting subregion within a more extended starbursting disk.
The complication in the latter case is that neighboring regions on the disk 
would unavoidably elevate some gas to varying altitudes, resulting in an
environment for the compact starbursting region in question that is subject to more obscuring gas,
in lines of sight deviated from the polar direction.
Nevertheless, we do expect that the LyC escape is, on average, an increasing function of 
the star formation surface density within an extended starburst,
unless ram-pressure becomes a dominant confining process, as likely in the case of most ULIRGs with respect
to the star formation rate surface density.

\section{Discussion and Conclusions}

A simple analysis of the dynamics of star forming clouds is performed.
The dynamical players include supernova driven outward momentum flux,
radiation pressure, gravitational force and ram pressure due to infalling gas.
The single most significant finding, evident in Figure (\ref{fig:sol3v}), 
is that galaxies able to promptly drive leftover gas away 
are characterized by two basic properties:
small sizes ($\le 1$kpc) 
and high star formation rate surface densities ($\Sigma_{\rm SFR}\ge 10\msun~{\rm yr}^{-1}~{\rm kpc}^{-2}$).
These characteristics are dictated by 
the twin requirements for removing obscurating material promptly: expelling the gas to a high altitude of the size of the system itself 
within a few million years since the end of the starburst (which coincides with the onset of the liftoff of the leftover gas from star formation).

As a matter of fact, the only physical commonality among the distinct classes of observed galaxies 
known to be LyC leakers
- green-peas galaxies, [SII]-weak galaxies, some LAEs -
are their high star formation rate surface densities and compact sizes.
Our analysis now provides a unifying physical origin for LyC leakers.
On the other hand, some other observed properties that differ among different classes 
are merely symptoms and consequences as a result of gas expulsion,
such as those related to density-bound structures and their manifestations
due to much reduced gas density around star formation regions
in the form of line emission and absorption by a relatively modest amount of gas
along the line of sight
(OIII emission, weak [SII] line, strong Ly$\alpha$ emission, etc).
The compactness of the starburst region likely requires some triggering event to drive
the gas to the central region of a new galaxy or an old galaxy rejuvenated.
Hence, looking for signs of significant gravitational interactions,
such as nearby companions or post merger features
will shed useful light.

In light of this clarification of the physical origin of prodigious LyC leakers,
a more robust way to search and identify LyC leakers may be to 
focus on their basic physical properties of compactness and high SFR surface density,
in addition to likely symptoms as a result of gas expulsion.
While such an approach will unite the various disparate kinds of observed LyC leakers physically,
it will also help broaden the range of methods that may be employed to search for LyC leakers,
which may be important for an adequate account of the overall abundance of LyC leakers.
For this undertaking, it is useful to highlight three other verifiable predictions for LyC leakers
 from this analysis:
(1) the newly formed stellar masses are in the range of $10^8-10^{10}\msun$,
(2) the outflow velocities are in the range typically of
$100-600\kms$, 
(3) there a positive correlation between the stellar masses formed and the outflow velocities.

Furthermore, two interesting by-product predictions are also borne out.
First, ULIRGs appear to be in a parameter region where they should be prodigious LyC leakers,
unless there is a large ram-pressure due to infalling gas with a rate exceeding about 30 times
the star formation rate.
Then, unavoidably, towards the tail end of a ULIRG event when the ram-pressure relents, 
advanced ULIRGs may turn significant LyC leakers.
Second, Lyman break galaxies with size exceeding $1$kpc are shown not to be prodigious LyC leakers.
Thus, if LBGs have significant LyC leakage, as latest observations appear to suggest,
it may be that the effective radii of their star forming regions have been over-estimated
by a factor of $2-4$.

Finally, an important physical trend is noted.
In a hierarchical structure formation model,
galaxies at high redshift are, as a whole,
more compact and have higher star formation rate surface densities,
simply due to the fact of the universal expansion.
In addition, more ubiquitous interactions among galaxies help drive low angular momentum
to the central regions of galaxies to facilitate formation of compact systems, on top of
already relatively smaller physical sizes of galaxies at high redshift.
Therefore, given what is learned in this analysis,
it may be predicted that the LyC escape fraction of galaxies
is expected to increase with increasing redshift
at a given star formation rate, 
a given stellar mass, a given luminosity or a given physical size of galaxies. 
We can also predict that the overall escape fraction for galaxy population as a whole
is expected to increase with increasing redshift.
This trend helps understand why galaxies at EoR may have
much higher escape fractions than their lower redshift counterparts
\citep[e.g.,][]{2009Wise, 2014Kimm}
and help provide the physical basis for stellar reionization.

We note in passing that
simulations that are not able to at least fully resolve star formation regions of size of $100$pc or so
may yield results significantly removed from reality, 
exacerbating plaguing issues, such as overcooling, over-metal enrichment, under-prediction of LyC photon, etc.

\vskip 1cm

I would like to thank Tim Heckman and Masami Ouchi for discussion,
Bengjie Wang and Tim Heckman for sharing observational data prior to publication,
and the warm hospitality of IPMU where this work was initiated.
The research is supported in part by NASA grant 80NSSC18K1101.


\end{document}